# The metal line systems in HS1700+6416: Evidence for inhomogeneities*

Patrick Petitjean[1,2], Rüdiger Riediger[3], and Michael Rauch[4]

[1] Institut d'Astrophysique de Paris - CNRS, 98bis Boulevard Arago, F-75014 Paris, France
[2] UA CNRS 173- DAEC, Observatoire de Paris-Meudon, F-92195 Meudon Principal Cedex, France
[3] Astrophysikalisches Institut Potsdam, An der Sternwarte 16, D-14482 Potsdam, Germany
[4] Carnegie Observatories, 813 Santa Barbara Street, Pasadena, CA 91101, USA

**Abstract.** We present a high S/N ratio optical spectrum of the bright quasar HS 1700+6416. These data usefully complement the UV HST data from Vogel & Reimers (1995). We analyse the metal line systems using photo-ionization models allowing for inhomogeneities in the gas. The models are able to reproduce within a factor of two the large $N(\text{He I})/N(\text{H I})$ ratio together with the mean column densities of the heavy element species observed in the $z \sim 2$ systems. The density contrast between low and high density regions is of the order of 30. Although the [O/C] abundance ratio seems slighly larger than solar, firm conclusion should await higher spectral resolution data. A break at the He II ionization limit of a factor of ten is acceptable in the ionizing spectrum. Abundances are found to be about 0.08 $Z_\odot$.

This together with other determinations from the literature indicates that there is evolution in the metal content of metal line systems with redshift: [Mg/H] is slightly below solar at $z \sim 0.7$ and [C/H] $\sim 10^{-2}$-$10^{-1}$ at $z \sim 2$.

We detect an associated C IV system at $z_{\text{abs}} = 2.7126$ in which there is some evidence for the presence of Ne VIII, probably Ne VI, Ne VII, and possibly Si XII.

## 1. Introduction

The bright high-redshift ($V = 16.1$, $z_{\text{em}} = 2.7$) QSO HS 1700+6416 (Reimers et al. 1989) has the particularity to contain 15 metal line systems with HI column densities in the range $2.5\ 10^{15}$-$1.1\ 10^{17}$ cm$^{-2}$ (Vogel & Reimers 1995 hereafter VR95). All these systems have small optical depth at the Lyman limit ($0.02 < \tau_{\text{LL}} < 1$) with the consequence that the UV flux shortward the redshifted Lyman limit is never completely absorbed. This situation makes HST spectroscopy possible and offers, for the first time, the opportunity to observe redshifted far-UV absorption lines of most of the C, N, O ions (Reimers et al. 1992, Reimers & Vogel 1993, Vogel & Reimers 1993, VR95).

Since most of the ionization stages of C, N and O are observed, it is possible to make detailed analysis of the systems to derive constraints on the physical state of the gas and to discuss abundances and how the gas is ionized. Vogel & Reimers (1993) have shown, in the framework of photo-ionization models assuming constant density, that carbon is underabundant relative to solar, [C/H] $\sim$ -2, and oxygen and nitrogen are overabundant compared to carbon by a factor of 3 to 8. They conclude from their analysis that the ionizing spectrum has no break at the He II ionization potential (228 Å). Reimers & Vogel (1993) are however unable to reproduce the large neutral helium to neutral hydrogen column density ratio ($N(\text{He I})/N(\text{H I}) \sim 0.03$). The latter could be explained in the framework of the decaying neutrino theory (Sciama 1994). However the subject is somewhat controversial (e.g. Miralda-Escudé & Ostriker 1992). Giroux et al. (1994) have argued that the observations are incompatible with photo-ionization equilibrium by a single meta-galactic background and can be explained if the absorbers possess a multiphase medium with hot gas in collisional equilibrium.

It may be possible however to reconcile pure photo-ionization models with the data. Indeed the assumption of constant density all over the absorber is highly questionable. It is well known from high spectral resolution data (e.g. Blades 1988) that metal line systems do not arise in an homogeneous medium. It has been shown that this has important consequences when interpreting the ionization structure of the gas (Petitjean et al. 1992).

In this paper we investigate in detail this possibility using complementary new optical data presented in Section 2. Photo-ionization models and their consequences are discussed in Section 3 and we draw our conclusions in Section 4.

## 2. The data

Optical echelle spectra of HS 1700+6416 have been obtained with the red channel spectrograph at the Multiple Mirror Telescope on Mt. Hopkins, Arizona, using a 150/mm echelette grating. The slit width was chosen to be 1 arcsec wide, for a resolution of FWHM $\sim 95$ km s$^{-1}$. A slit length of 20 arcseconds provided enough sky background to allow proper sky subtraction even in the sky emission line region. The data consist of two exposures of 3600 sec duration taken under somewhat cloudy conditions in September 1994. Fig. 1 shows the resulting normalized spectrum in the wavelength range 455–820 nm. Outside regions of atmospheric absorption and emission, the S/N ratio is in the range 30-45. Equivalent widths and heliocentric wavelengths for identified lines are given in Table 1.

### 2.1. Comments on individual systems

#### 2.1.1. $z_{\text{abs}} = 0.7221$

We detect the two lines of the Mg II doublet. Mg II$\lambda$2796 is blended with Si IV$\lambda$1402 at $z_{\text{abs}} = 2.4334$. Using the equivalent width observed for Mg II$\lambda$2803 and the optically thin case we find a lower limit for the Mg II column density $N(\text{Mg II}) > 5\ 10^{12}$ cm$^{-2}$.

On the basis of the Ly$\alpha$, Ly$\beta$ and Ly$\gamma$ lines, VR95 derive $N(\text{H I}) = 1.6\ 10^{16}$ cm$^{-2}$ in this system. The H I column density cannot be much larger as there is no depression in the QSO continuum at the wavelength where the Lyman discontinuity is to be found (Reimers et al. 1993). Although the uncertainties are large, this shows that this system is another member of the population of $\tau_{\text{LL}} < 1$ Mg II systems detected by the HST (Bergeron et al. 1994). The latter authors have shown, using simple argumentation, that this is indicative of abundances close to solar. Using their Eq.(3), and the above lower limit for $N(\text{Mg II})$, we derive [Mg/H] $> -0.22$.

O VI$\lambda\lambda$1031,1037 lines could be present (see VR95). However using our redshift estimate, O VI$\lambda$1031 is shifted by more than 1 Å from the expected position in the VR95 data (to be compared to the 1.44 Å FWHM resolution). O VI$\lambda$1037 coincides with a weak feature detected only at the 3 $\sigma$ level. If the O VI absorption is real, it is clear that it is impossible to explain both Mg II and O VI absorptions assuming constant density in the absorbing cloud. The structure of the absorber is certainly more like a large halo of high excitation and low density, where O VI is produced, surrounding a region of higher density and lower excitation explaining the Mg II absorption (see Bergeron et al. 1994 and Riediger & Petitjean 1995).

#### 2.1.2. $z_{\text{abs}} = 0.8642$

We detect Mg II$\lambda$2796 at the $3\sigma$ level. If this feature is real, and considering the optically thin case which minimizes the Mg II column density, we obtain $N(\text{Mg II}) > 1.9\ 10^{12}$ cm$^{-2}$. Using the same arguments as in section 2.1.1 we derive a magnesium abundance [Mg/H] $> -0.77$.

The Ca II lines are redshifted in a part of the spectrum affected by atmospheric absorption but the observed equivalent width of both lines should be smaller than 0.2 Å. The Fe II$\lambda\lambda$2596,2600 doublet is not detected with $w_{\text{r}} < 0.04$ Å.

#### 2.1.3. $z_{\text{abs}} = 1.1573$

We unambiguously detect the Mg II doublet. Weak Fe II$\lambda\lambda$2586,2600 lines may be present; Fe II$\lambda$ 2374 is not detected ($w_{\text{r}} < 0.05$ Å) and Fe II$\lambda$2382 is blended with C IV$\lambda$1550 at $z_{\text{abs}} = 2.3155$.

The Doppler parameter is not well constrained as the doublet is almost optically thin. The column density is in the range $2.8\ 10^{12}$-$1.6\ 10^{13}$ cm$^{-2}$. From the size of the Ly$\alpha$ edge, VR95 derive $N(\text{H I}) = 7.1\ 10^{16}$ cm$^{-2}$. Using the same arguments as in Section 2.1.1, we find [Mg/H] $> -1.1$ which is compatible with the findings of VR95 ([C/H] = –0.44).

#### 2.1.4. $z_{\text{abs}} = 1.4735$

This system is considered as uncertain by VR95. We detect neither Mg II$\lambda$2796 nor Fe II$\lambda\lambda$2586,2600 with $w_{\text{r}} < 0.025$ Å and conclude that the system is certainly not real.

#### 2.1.5. $z_{\text{abs}} = 1.7242$

A strong feature, $w_{\text{obs}} = 1.2$ Å, is detected at $\lambda$4551.6 corresponding to Al II at $z_{\text{abs}} = 1.7242$. The corresponding Fe II and Al III lines which are redshifted in a very good part of the spectrum are not detected, Fe II$\lambda\lambda$2586,2600 and Al III$\lambda$1854 have $w_{\text{r}} < 0.01$ and 0.02 Å respectively.

If the Al II identification is correct, the Al II column density is larger than $9\ 10^{12}$ cm$^{-2}$. The H I column density derived from the Lyman limit is $1\ 10^{17}$ cm$^{-2}$ (VR95). Using the models by Petitjean et al. (1994), we derive an aluminium abundance larger than one tenth of the solar value. The non detection of Fe II is not a problem and would indicate a somewhat large ionization factor.

Alternatively the line could be a H I Ly$\alpha$ line at $z = 2.744$.

**Table 1.** Identified absorption lines

| $\lambda_{\text{helio}}$ | $w_{\text{obs}}$ | $\sigma^a$ | Identification | $z_{\text{abs}}$ |
|---|---|---|---|---|
| 4551.61 | 1.22 | 0.15 | Al II$\lambda$1670: | 1.7242 |
| 4598.96 | 0.56 | 0.05 | N V$\lambda$1238 | 2.7124 |
| 4613.55 | 0.54 | 0.05 | N V$\lambda$1242 | 2.7122 |
| 4620.55 | 1.32 | 0.05 | Si IV$\lambda$1393 | 2.3152 |
| 4650.47 | 0.77 | 0.05 | Si IV$\lambda$1402 | 2.3152 |
| 4785.38 | 0.28 | 0.07 | Si IV$\lambda$1393 | 2.4334 |
| 4815.71 | 0.34 | 0.04 | Mg II$\lambda$2796 | 0.7221 |
|         |      |      | Si IV$\lambda$1402 | 2.4334 |
| 4828.01 | 0.20 | 0.04 | Mg II$\lambda$2803 | 0.7221 |
| 4904.52 | 0.56 | 0.03 | C IV$\lambda$1548 | 2.1679 |
| 4913.10 | 0.35 | 0.05 | C IV$\lambda$1550 | 2.1682 |
| 5061.36 | 0.11 | 0.03 | Si II$\lambda$1526 | 2.3152 |
| 5121.69 | 0.45 | 0.04 | C IV$\lambda$1548 | 2.3082 |
| 5133.06 | 3.00 | 0.05 | C IV$\lambda$1548 | 2.3155 |
|         |      |      | C IV$\lambda$1550 | 2.3082 |
| 5141.64 | 2.32 | 0.05 | C IV$\lambda$1550 | 2.3156 |
| 5214.16 | 0.15 | 0.04 | Mg II$\lambda$2796 | 0.8646 |
| 5232.94 | 0.11 | 0.05 | Mg II$\lambda$2803 | 0.8666 |
| 5314.96 | 0.42 | 0.04 | C IV$\lambda$1548 | 2.4330 |
| 5324.34 | 0.83 | 0.04 | C IV$\lambda$1550 | 2.4330 |
|         |      |      | C IV$\lambda$1548 | 2.4392 |
| 5333.42 | 0.31 | 0.04 | C IV$\lambda$1550 | 2.4392 |
| 5539.94 | 0.39 | 0.03 | C IV$\lambda$1548 | 2.5783 |
| 5549.55 | 0.16 | 0.03 | C IV$\lambda$1550 | 2.5786 |
| 5579.17 | 0.19 | 0.04 | Fe II$\lambda$2586 | 1.1569 |
| 5608.68 | 0.11 | 0.04 | Fe II$\lambda$2600 | 1.1570 |
| 5747.63 | 0.33 | 0.03 | C IV$\lambda$1548 | 2.7125 |
| 5757.17 | 0.19 | 0.03 | C IV$\lambda$1550 | 2.7126 |
| 6032.47 | 0.26 | 0.04 | Mg II$\lambda$2796 | 1.1573 |
| 6048.07 | 0.17 | 0.05 | Mg II$\lambda$2803 | 1.1573 |

$^a$ noise rms in the adjacent continuum

### 2.1.6. $z_{\text{abs}} = 1.8465$

We detect neither Mg II$\lambda$2796 nor Fe II$\lambda\lambda$2586,2600 ($w_r < 0.04$ and 0.02 Å respectively). This system is ascertained by a Lyman edge however and a few metal lines in the UV (VR95). The absence of Fe II and Mg II is consistent with the non detection of C II lines in the UV.

### 2.1.7. $z_{\text{abs}} = 2.1678$

Strong C IV absorption is present in this system with $w_r(\lambda\lambda1548,1550) = 0.18$ and 0.11 Å for the two lines. The equivalent width for the former line is twice as large as that given by VR95. None of the Fe II and Al II lines is detected, Fe II$\lambda\lambda$2586,2600 and Al II$\lambda$1670 are $w_r < 0.02$ and 0.03 Å respectively. This is consistent with the non detection of Si II and Si III lines in the HST data.

From the doublet ratio, the column density is in the range 5 - 8 $10^{13}$ cm$^{-2}$. The Doppler parameter is not well constrained, $8 < b < 19$ km s$^{-1}$.

### 2.1.8. $z_{\text{abs}} = 2.189$ and 2.290

Although strong Ly$\alpha$ lines are present at both redshifts, the presence of metals is far from certain. The only good UV coincidences are C III$\lambda$977 for the former and O V$\lambda$629 for the latter. Neither N III nor O IV is detected at $z_{\text{abs}} = 2.189$ (VR95).

We do not detect C IV in our data with $w_r < 0.02$ Å, corresponding to $N(\text{C IV}) < 5\ 10^{12}$ cm$^{-2}$ in the optically thin case.

### 2.1.9. $z_{\text{abs}} = 2.3082$ and 2.3155

These systems are separated by about 600 km s$^{-1}$. A strong C IV absorption is associated with the latter, $w_r(\text{C IV}\lambda\lambda1548,1550) = 0.90$ and 0.70 Å respectively. This is about four times larger than the value reported by VR95. This is obviously a blend of several components and the two systems could be physically associated. In that case, the complex is part of the subpopulation of C IV systems which are spread over large velocity ranges, $\Delta V > 500$ km s$^{-1}$, and have comparatively small equivalent widths, $w_r < 1$ Å (Petitjean & Bergeron 1994). Strong Si IV absorption and weak Si II$\lambda$1526 ($w_r = 33$ mÅ) are seen in the $z_{\text{abs}} = 2.3155$ system. Neither Al II$\lambda$1670 nor Al III$\lambda\lambda$1842,1854 is detected at a level $w_r < 0.05$ Å.

### 2.1.10. $z_{\text{abs}} = 2.4330$ and 2.4392

The two systems are separated by about 524 km s$^{-1}$ and have broad and irregular C IV absorptions revealing, even at our resolution, complex structures. A 2$\sigma$ feature with $w_{\text{obs}} = 0.08$ Å is present at the position of Si II$\lambda$1526 at $z_{\text{abs}} = 2.4330$; in the latter system, a weak Si IV$\lambda$1393 line is present. Nothing is seen for Al II and Al III lines with $w_{\text{obs}} < 0.03$ Å. Identification of lines in the UV is difficult as most of them are part of complex blends. However Reimers et al. (1995) confirm the detection of O II$\lambda\lambda\lambda$376,430,392 at $z_{\text{abs}} = 2.433$ from GHRS observations. The present Si II and Al II observations (corresponding to $N(\text{Al II}) < 2\ 10^{11}$ cm$^{-2}$ and $N(\text{Si II}) \sim 5\ 10^{11}$ cm$^{-2}$), could be difficult to explain if the O II column density is found larger than $5\ 10^{13}$ cm$^{-2}$.

### 2.1.11. $z_{\text{abs}} = 2.5783$

Weak C IV absorption is detected at this redshift with $w_r(\text{C IV}\lambda\lambda1548,1550) = 0.11, 0.05$ Å.

### 2.1.12. $z_{\text{abs}} = 2.7126$

We detect an associated system with N V stronger than C IV. This system was not identified on the basis of HST

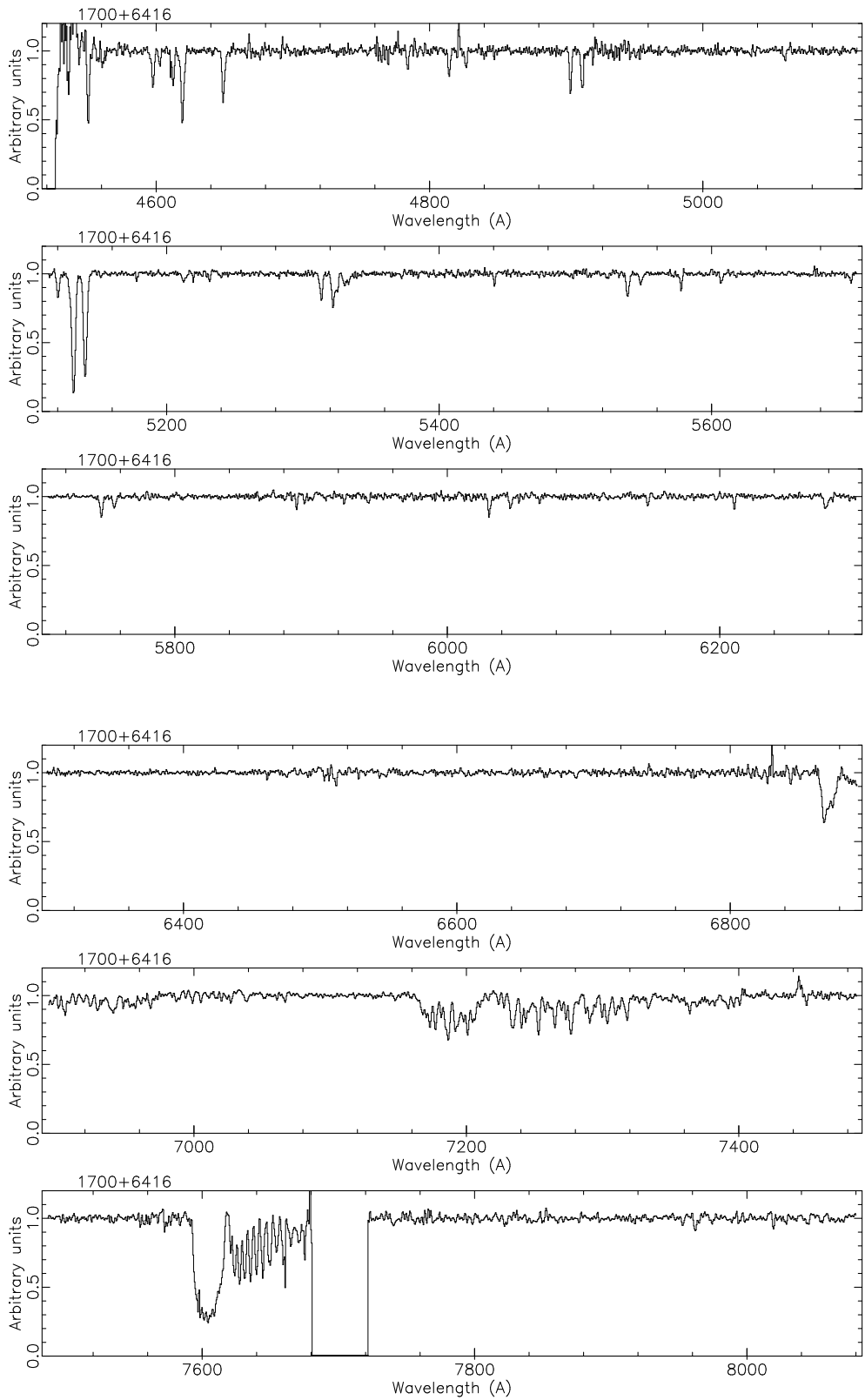

**Fig. 1.** Normalized spectrum of HS 1700+6416

data only. It is interesting however to go back to the data to attempt identifying lines using the optical redshift.

Ne V$\lambda$568 could be present in the wing of C III$\lambda$977 at $z_{\text{abs}} = 2.29$. A strong absorption feature is seen at the position expected for Ne VII$\lambda$465. Part of it should be H I$\lambda$926 at $z_{\text{abs}} = 0.8642$ but the fit is unsatisfactory and allows for additional absorption. More convincing is Ne VIII$\lambda$770 for which a strong detached absorption feature is present. The latter was identified as O II$\lambda$833 at $z_{\text{abs}} = 2.433$. The fit is not satisfactory however and the O II column density may have been overestimated (see Section 2.1.10). We thus believe that Ne VIII is present in this system. The second line of the doublet, Ne VIII$\lambda$780, falls in a strong blend but a narrow feature is observed at the expected wavelength. Ne VI$\lambda$558 could be present as well, as a feature is seen at the expected wavelength which was identified as O V$\lambda$629 at $z_{\text{abs}} = 2.29$. The latter identification is doubtful since no absorption is seen from O IV$\lambda\lambda$608,787 (see Section 2.1.8). O V$\lambda$629 is seen as a weak line. This identification would cast some doubt on the N II$\lambda$1083 line at $z_{\text{abs}} = 1.1572$ which is displaced in wavelength anyway and too strong in the VR95 fit. The redshifted Ar V$\lambda$822 line falls in a very broad feature but is certainly weak; there is a weak feature at the position of Ar VI$\lambda$767 and Ar VI$\lambda$755 (Wiese et al. 1969) could be present in the wing of Mg II$\lambda$2803 at $z_{\text{abs}} = 0$. Strong narrow features are seen at the positions of Ar VIII$\lambda\lambda$700,713; other probable identifications are given by VR95 but the fit allows for additional absorption especially for the stronger line. More surprising, there is an isolated feature at $\lambda$1853.5 which is exactly at the expected wavelength for Si XII$\lambda$499 (Verner et al. 1994). The feature has been identified as Al III$\lambda$1854 by VR95. However the wavelength coincidence is not perfect (see their fit) and Al III$\lambda$1862 is definitively not present. If the Si XII identification is correct, either photons with energy larger than 476 eV are required or the gas is collisionally ionized and the corresponding temperature should then be larger than $1.9 \, 10^6$ K. Given the very high ionization state of this system, the associated He I lines are not expected.

Possible detection of Ne VIII$\lambda$770 has already been reported at $z_{\text{abs}} = 2.1373$ towards UM675 ($z_{\text{em}} = 2.148$, Burbidge 1992) and from the BAL system in 0226-1024 (Korista et al. 1992). The HS 1700+6416 observations are qualitatively consistent with absorption increasing with the ionization potential: smaller than 70 eV for C IV and Ar V, in the range 70-100 eV for N V, O V and Ar VI and larger than 100 eV for Ar VIII, Ne VII and Ne VIII. In the case of photo-ionization, the ionization parameter must be very high and photons with energy larger than 100-250 eV must substantially contribute to the ionization. In the absence of H I column density determination it would be difficult to go further and infer any metal abundances.

## 3. Interpretation

Vogel & Reimers (1993), Reimers and Vogel (1993) and VR95 have analysed the column densities in the heavy-element systems using photoionization models with *constant density*. They find evidence for an enhanced oxygen abundance compared to carbon in four of the systems ($z_{\text{abs}} = 1.8465, 2.1678, 2.308$ and $2.433$). They also find that the models of the $z_{\text{abs}} = 1.8465$ and $2.1678$ systems exclude a 228 Å break of more than a factor of 5 in the ionizing radiation spectrum. They argue that in a few of the systems (especially $z_{\text{abs}} = 2.1678$ and $2.433$), the $N(\text{He I})/N(\text{H I})$ column density ratio is a factor of 5 too large to be explained by the above photo-ionization models. We shall make a few remarks which should help resolving these contradictions in the framework of photo-ionization. To illustrate the discussion we have defined a standard system taking typical column densities as given by VR95. The latter are given in column 2 of Table 2.

### 3.1. The $N(He\,\textsc{i})/N(H\,\textsc{i})$ ratio

Although some doubt can be cast on some of the O VI detections (see Section 3.2), the O V absorptions are clearly present and strong in most of the systems toward HS1700+6416. In order to explain the large O V and O VI column densities, one has to maximize the O V/O and O VI/O ratios. This implies a high ionization parameter (ratio of the density of ionizing photons to the hydrogen density) of the order of $U \sim 2 \, 10^{-3}$ which leads to a density of $3.5 \, 10^{-3}$ particle per cm$^{-3}$ for a typical flux of $5 \, 10^{-22}$ erg cm$^{-2}$ s$^{-1}$ Hz$^{-1}$ sr$^{-1}$ at the hydrogen Lyman limit. In such a model, H I/He I $\sim$ 175 that is 7 times larger than observed.

One way to reconcile photo-ionization models with observations is to consider that highly ionized species, such as O V or O VI, and neutral elements originate in regions of different density. Indeed, when observed at high spectral resolution, metal lines split into several components spanning typically a few hundreds km s$^{-1}$ (e.g. Petitjean & Bergeron 1994). This suggests that the absorbing medium is inhomogeneous and several clouds of different ionization stages are present along the line of sight.

Models of inhomogeneous clouds have been described by Petitjean et al. (1992). These authors have shown that such models explain well the statistical properties of metal line systems at redshifts 2-2.5. A two-phase model was necessary as well to explain consistently absorptions observed in the $z_{\text{abs}} = 0.791$ system toward PKS 2145+06 (Bergeron et al. 1994). O VI and N V originate in the low density phase, the neutral and singly ionized species are found in the high density phase, and both phases can contribute to the C IV column density. In such models H I/He I $\sim$ 20 (see Table 5 of Bergeron et al. 1994) which matches the value observed in HS 1700+6416.

We therefore conclude that the large observed $N(\text{He I})/N(\text{H I})$ ratio is indicative of the presence of inhomogeneities in the absorber.

### 3.2. The OVI phase

Six of the systems described in the previous section have been claimed to have associated O VI absorption. Although coincidence with Ly$\alpha$ lines cannot be excluded, there is good evidence for the presence of both lines of the doublet at $z_{\text{abs}} = 0.7217$ and $1.1572$. The other O VI detections are not secure: in the $z_{\text{abs}} = 1.725$ system, O VI$\lambda 1037$ is certainly not present and O VI$\lambda 1031$ is lost in a strong blend; in the $z_{\text{abs}} = 2.1678$ system, only O VI$\lambda 1031$ is within the observed wavelength range and it falls in a strong blend. For the other systems, the lines are shifted in the optical wavelength range for which better data are definitively needed (see Section 2.1.9 and VR95).

The O VI doublet has been detected in four of the five $z_{\text{abs}} < z_{\text{em}}$ systems observed at $0.6 < z < 1.1$ by the HST (Bergeron et al. 1994). It is seen also in a composite spectrum formed from a large number of C IV systems with velocity difference from the emission redshift larger than 5000 km s$^{-1}$ (Lu & Savage 1993). However associated systems with abundances in excess of solar have been found up to 10000 km s$^{-1}$ from the emission redshift (Petitjean et al. 1994) and it is unclear whether the O VI absorption in the composite spectrum could be due to associated systems (see Savaglio et al. 1994).

Therefore, if there is no doubt that strong O VI absorption is common in $z \sim 1$ systems, its presence is still to be confirmed at higher redshift.

### 3.3. Ionization structure

We have taken, as a basis for discussion, column densities from Table 4 of VR95, characteristic of the high redshift systems (column 2 of Table 2). We use a modified version of the code Nebula (Péquignot et al. 1978, Petitjean et al. 1990). The absorbing cloud is modelled as a slab photo-ionized by an external isotropic flux $F_\nu$. The density through the slab is *not* constant. Basically there are two regions of constant density, the central part of the cloud having a density $f$ times larger than the outer region. The photo-ionization code cannot handle a discontinuity between both regions. To avoid difficulties (and mimic most probably better the real situation), the density decreases continuously from the inner to the outer regions within a distance equal to 20% of the inner region dimension. Abundances are taken a fraction of the solar values with the latter from Gehren (1988). The photoionizing spectrum is a power law of index $\alpha = -0.6$ with a break of a factor of 10 at the He II ionization limit (Madau 1992). Results from our best model are given in column 3 of Table 2; contributions of the highly ionized phase (HIP; low density region) and low ionization phase (LIP; high density region) are given in columns 4 and 5 respectively. For a flux at the hydrogen Lyman limit $F_{\text{o}} = 5\ 10^{-22}$ erg s$^{-1}$cm$^{-2}$Hz$^{-1}$sr$^{-1}$ and abundances 0.08 Z$_\odot$, the density and dimension of the HIP and LIP are 4 10$^{-4}$ cm$^{-3}$, 25 kpc and 0.036 cm$^{-3}$, 75 pc respectively.

**Table 2.** Column densities

| Species | Obs[a] | Model | HIP[b] | LIP[c] |
|---|---|---|---|---|
| HI | 5.5(16) | 3.2(16) | 4.8(14) | 3.2(16) |
| HeI | 2.2(15) | 2.2(15) | 1.3(12) | 2.2(15) |
| HeII |  | 3.5(17) | 1.9(16) | 3.3(17) |
| CII | 5.0(13) | 1.9(13) | 5.8(10) | 1.9(13) |
| CIII | 1.2(14) | 1.3(14) | 1.5(13) | 1.2(14) |
| CIV | 5.7(13) | 1.3(14) | 1.2(14) | 1.1(13) |
| CV |  | 6.6(14) | 6.6(14) | 2.1(12) |
| NII |  | 6.4(12) |  | 6.4(12) |
| NIII | 5.0(13) | 3.5(13) | 4.6(12) | 3.0(13) |
| NIV | 1.0(14) | 6.9(13) | 6.5(13) | 3.9(12) |
| NV | 5.0(13) | 7.7(13) | 7.7(13) | 2.9(11) |
| OII | 1.0(14) | 4.0(13) | 2.7(10) | 4.0(13) |
| OIII | 3.2(14) | 2.6(14) | 3.4(13) | 2.3(14) |
| OIV | 8.2(14) | 4.4(14) | 4.2(14) | 1.7(13) |
| OV | 1.0(15) | 7.7(14) | 7.7(14) | 2.8(12) |
| OVI | 3.5(14) | 5.8(14) | 5.8(14) |  |
| SiII | 1.0(13) | 1.7(12) |  | 1.7(12) |
| SiIII | 1.0(13) | 9.1(12) |  | 9.1(12) |
| SiIV |  | 1.9(12) |  | 1.9(12) |

[a] see VR95; 1.0(14) means 1.0 $10^{14}$ cm$^{-2}$
[b] High ionization phase see text
[c] Low ionization phase

It is apparent from Table 2 that the model fits well the observations within a factor of two which can be considered as a lower limit for uncertainties in most of the column density determinations. The [C/O] abundance ratio seems slightly small. However it is unclear whether this can be considered as an unambiguous result since the C IV column density is very uncertain (see Section 2.).

The fit could be better if the observed O V and O VI column densities were overestimated. In the previous Section we have seen that it is unclear whether or not O VI absorption is present in systems at $z_{\text{abs}} > 2$. On the contrary it is clear that in all the systems O V absorption is strong. This does not imply however that the O V column density is large.

To a good approximation the lines are build up of a number, $n$, of subcomponents of similar strength, $w_{\text{o}} = w_{\text{r}}/n$, and similar Doppler parameter, $b_{\text{o}}$, as indicated by the correlation between the equivalent width and the number of components seen at high spectral resolution (Petitjean & Bergeron 1990, 1994). If the line is not strongly saturated, the Doppler parameter measured from low resolution data, is indicative of the number of components, $b = nb_{\text{o}}$. The total column density can be written,

$N_\text{tot} = nN(b_\text{o}, w_\text{o})$. This means that the column density measured from low resolution data is a good estimate of the true column density if the lines are not strongly saturated and if the $b$ value used is large enough so that it is representative of the number of subcomponents.

It is possible that the H I and O V absorptions do not arise in the same regions. In our picture O V arises in a more dilute phase than H I. The number of subcomponents and the spread in velocity of the subcomponents in the O V absorption are certainly larger than for low excitation species. This is observed in systems for which high and low excitation lines are both detected (e.g. Turnshek et al. 1989). Therefore the $b$ values used to derive $N$ from $w$ should be larger for O V than for H I. Consequently it is very important to obtain good high spectral resolution data in the optical to separate the subcomponents of the different systems and then use this information to better fit the UV lines.

## 4. Conclusion

We have analysed the absorption line spectrum of HS 1700+6416 in the light of new optical data. We show that the column density ratios [including $N(\text{He I})/N(\text{H I})$] are consistent with the presence of inhomogeneities in the gas with density variations of the order of 10 to 100. We do not find compelling evidence for systematic [O/C] overabundance as compared to the solar values.

We detect an associated C IV system at $z_\text{abs} = 2.7126$ in which there is some evidence for the presence of Ne VIII, probably Ne VI, Ne VII, and possibly Si XII.

We derive abundances $Z = 0.2$ to $0.6\ Z_\odot$ in two $z < 1$ ($z_\text{abs} = 0.7221, 0.8642$) systems. These values are similar to that of $Z = 0.4\ Z_\odot$ derived in the $z_\text{abs} = 0.7913$ towards PKS 2145+06 by Bergeron et al. (1994). We find that abundances are of the order of $Z = 0.08\ Z_\odot$ for $z \sim 2$ systems. Similar values have been derived in four LLS systems towards PKS 0424–131 and Q 0450–131 (Petitjean et al. 1994). At higher redshift $z \sim 3$, abundances in LLSs may be of the order of $10^{-2}\ Z_\odot$ or less (Steidel 1990).

Abundance determinations are performed in LLS or damped systems. The latter are well suited to this purpose since the H I column density is well determined, the ionization correction is negligible and weak lines from Zn II and Cr II can be searched for (Meyer & York 1987). The results spread over several orders of magnitude, with abundances relative to the solar values in the range $[\text{Zn/H}] = 10^{-3}$-1 (Pettini et al. 1994). This may not be surprising since these systems can certainly arise through disks, in the proximity of star forming regions, and halos of galaxies (Petitjean et al. 1992). In this respect metal line systems with Lyman limit optical depth close to unity are useful to study the overall evolution of the metal content of the universe since they may arise in more homogenized gas. Abundances determination in these systems requires to model the ionization state of the gas and a large amount of information must be gathered to constrain the models. It seems however that abundances in metal line systems are of the order of $10^{-2}, 10^{-1}, 0.5\ Z_\odot$ at redshift 3 (Steidel 1990), 2 (Petitjean et al. 1992, this paper) and 0.5 (Bergeron et al. 1994, this paper) respectively. There is thus an indication for evolution of the metal content of metal line systems with redshift.